\documentclass[10pt,conference]{IEEEtran}
\IEEEoverridecommandlockouts
\usepackage{cite}
\usepackage{amsmath,amssymb,amsfonts}
\usepackage{algorithmic}
\usepackage{graphicx}
\usepackage{textcomp}
\usepackage{xcolor}
\usepackage{hyperref}
\def\BibTeX{{\rm B\kern-.05em{\sc i\kern-.025em b}\kern-.08em
    T\kern-.1667em\lower.7ex\hbox{E}\kern-.125emX}}
\usepackage{multirow}
\begin{document}
\title{Automated Bug Report Prioritization in Large Open-Source Projects}

\author{
\IEEEauthorblockN{Riley Pierson}
\IEEEauthorblockA{\textit{Department of Computer Science} \\
\textit{University of Colorado Colorado Springs}\\
\textit{and Siena College, Loudonville, NY, United States}\\
ra26pier@siena.edu} ~\\
\and
\IEEEauthorblockN{Armin Moin}
\IEEEauthorblockA{\textit{Department of Computer Science} \\
\textit{University of Colorado Colorado Springs}\\
\textit{United States} \\
amoin@uccs.edu \\
}
}

\maketitle

\begin{abstract}
Large open-source projects receive a large number of issues (known as \textit{bugs}), including software defect (i.e., bug) reports and new feature requests from their user and developer communities at a fast rate. The often limited project resources do not allow them to deal with all issues. Instead, they have to prioritize them according to the project's priorities and the issues' severities. In this paper, we propose a novel approach to automated bug prioritization based on the natural language text of the bug reports that are stored in the open bug repositories of the issue-tracking systems. We conduct topic modeling using a variant of LDA called TopicMiner-MTM and text classification with the BERT large language model to achieve a higher performance level compared to the state-of-the-art. Experimental results using an existing reference dataset containing 85,156 bug reports of the Eclipse Platform project indicate that we outperform existing approaches in terms of Accuracy, Precision, Recall, and F1-measure of the bug report priority prediction.
\end{abstract}


\section{Introduction}\label{introduction}
Large open-source projects offer an issue-tracking system with an open bug repository, where developers and users can report the software defects they find or any new feature requests they may have. These reports are called \textit{bug reports}. However, the projects' resources are limited, while processing and resolving the bug reports is typically very costly. Hence, not all bug reports in the open bug repository can be processed and handled at once. Since some of them are of higher importance/priority for the project, developers should resolve these bugs first. 

\textit{Bug triage} is a crucial process that excludes invalid and duplicate bug reports and assigns valid and unique ones to the developers working on the project so they can fix and resolve pertinent issues. In fact, part of the bug triage process is the task of priority assessment for each bug report based on the project's specific requirements, priorities, and available resources. In large open-source projects, a group of developers called \textit{triagers} are often responsible for the bug triage task. Over the past decades, (semi-)automated approaches based on various Artificial Intelligence (AI) methods and techniques have been proposed to make the bug triage process more effective and efficient. This way, the total project cost will be reduced under automated approaches since the triagers' and other developers' time will be saved and used more reasonably. 

In this paper, we focus on the automated priority assessment task in the bug triage process. A more effective and efficient priority assessment for bug reports will help developers put their time and energy into the most vital issues of the project. It is clear that an automated approach based on the history of bug resolution and software development in the project is only possible if the historical data in the bug repository have importance/priority tags or labels assigned to them, similar to what we see in Bugzilla and Jira, which are popular issue-tracking systems. If there is no such notion, for example, as is the case with GitHub and GitLab issues, which are also popular, our automated approach could not be applied.

In general, it is essential to distinguish between \textit{importance/priority} and \textit{severity}, although in some open-source systems, such as Android, they are used interchangeably \cite{Umer+2018}. Bug reporters (i.e., the users or developers reporting issues) usually assign severity levels to the respective issues, thus describing the extent of the bugs' impacts on them. In contrast, priorities are assigned by the developers who are processing issues (e.g., bug triagers). Priority levels characterize the importance they place from their perspectives on resolving the respective issues. In Bugzilla, which is the one we concentrate on in this paper, P1 is the highest importance/priority level, whereas P5 is the lowest \cite{WTP-Severity-Priority}.

As explained in Section \ref{related-work}, prior works in the literature have studied the bug triage problem and proposed various AI-based approaches to address it, such as approaches based on Machine Learning (ML) \cite{Anvik+2006} and Information Retrieval (IR) \cite{Sun+2011}. Most of them have used the natural language textual data in the bug reports as a crucial source of information. In some cases, temporal data, such as bug \textit{tossing} sequences, have also been taken into account \cite{Jeong+2009}.

This paper delivers a novel approach to automated priority assessment of bug reports in open bug repositories. First, we create numerous topic models and assign each report to one of them. This step is based on the work of Xia et al. \cite{Xia+2017}. Second, we train a text classifier for each topic, which can predict the priority level of a new bug report that belongs to that topic. For the latter, we deploy three different text classification methods, namely Gaussian Na\"ive Bayes, Multinomial Na\"ive Bayes, and Bidirectional Encoder Representations from Transformers (BERT). The latter method (i.e., BERT) is based on the work of Ali et al. \cite{Ali+2024}.

Our contribution is twofold: First, we propose and implement a novel approach to automated bug report priority prediction, validated by our experimental results using available data from an open reference dataset. Second, we provide an open-source prototype that enables other open-source projects using similar issue-tracking systems, which contain priority levels for issues, to use the proposed approach, thus benefiting from effective and efficient automated priority assessment.

This paper is structured as follows: Section \ref{background} provides some background information about open bug repositories of open-source projects and about the research field of Natural Language Processing (NLP). Further, we review the literature in Section \ref{related-work}. In Section \ref{proposed-approach}, we propose our novel approach and report on our experimental results in Section \ref{experimental-results}. Moreover, Section \ref{discussion} discusses the results and points out potential threats to validity. Finally, we conclude and suggest future work in Section \ref{conclusion-future-work}.

\section{Background}\label{background}

In this section, we provide some required background information about open bug repositories in open-source software projects and Natural Language Processing (NLP).

\subsection{\textbf{Open Bug Repositories}}

An \textit{issue-tracking system} is an essential part of a software development infrastructure. It serves as a \textit{ticket-tracking system} that enables tracking the status of issues, namely software defect reports and new feature requests. Each issue may be initiated by a software developer or a user of the project. In open-source software projects, the issue-tracking system is open to the public. Therefore, it is also known as an open bug repository. Various software solutions exist that provide such a service. For instance, Buzilla, JIRA, and GitHub/GitLab Issues are examples of well-established issue-tracking systems. Open bug repositories are very advantageous for open-source projects  \cite{Anvik+2005} and make the process of software evolution and maintenance transparent and more inclusive.

However, this comes at a price: Open-source projects have to dedicate significant resources to processing the incoming issues. Some of these issues are clearly invalid, unreproducible, or redundant. Others have to be prioritized and then assigned to developers so they can fix them. Nevertheless, in some cases, developers report issues they have detected and immediately assign them to themselves. 

\begin{figure*}[htb!]
    \centering
    \includegraphics[width=0.75\linewidth]{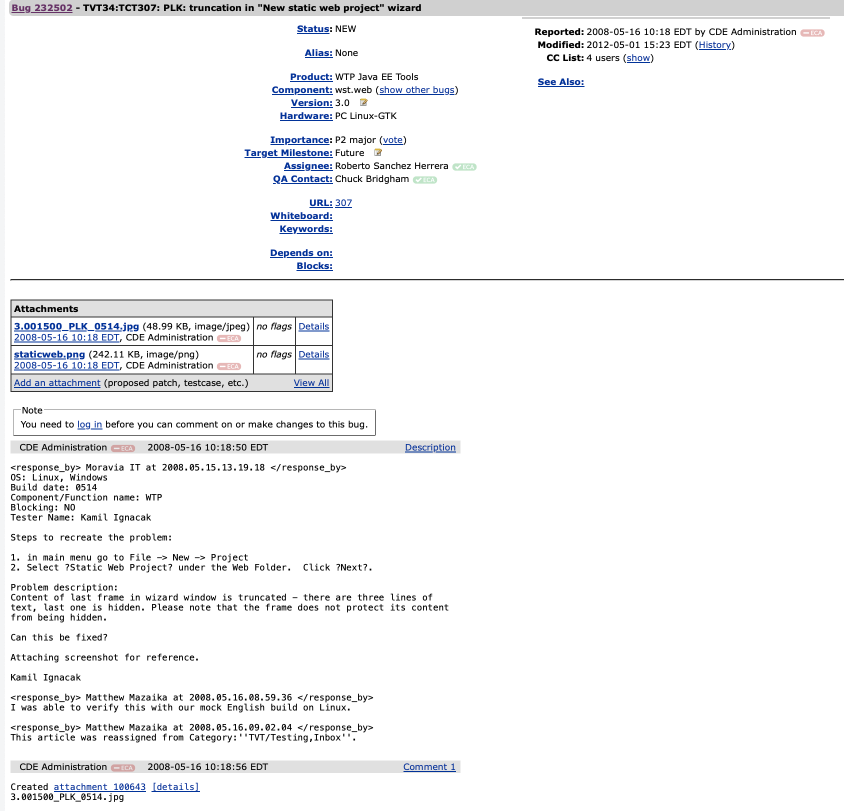}
    \caption{\centering{A shortened example of a Bugzilla bug report of the Eclipse WTP Java EE Tools project/product \cite{bugreport}}}
    \label{fig:Bugzilla-Example}
\end{figure*}

As mentioned in Section \ref{introduction}, in Bugzilla, bug reports are prioritized on a P1 through P5 scale. Developers consider P1 the most detrimental and impactful according to the project's goals and priorities, while P5 is reserved for the most manageable bugs within a specific product and component. Along with the prioritization level, each bug report includes some essential elements, including a unique bug ID, summary (title), status, product, component, reporter, assignee, and description. Summary and description are free-form text fields. Further, other developers or users may add additional comments (also free-form text) to the description. In contrast, product and component are examples of textual data with pre-defined sets of choices. Figure \ref{fig:Bugzilla-Example} illustrates part of a Bugzilla bug report in the Eclipse bug repository \cite{bugreport}.

The \textit{bug id} (e.g., 232502 in Figure \ref{fig:Bugzilla-Example}) is a unique identifier for an issue. Moreover, the \textit{summary (title)} \lq{}TVT34...\rq{} is something that the bug reporter has picked for this issue. Also, the \textit{status} field indicates the latest status of this issue during its life cycle. For instance, \textit{UNCONFIRMED} is used for issues that have not been reproduced by other developers/users or on a different system, whereas \textit{NEW} is assigned to issues that can be reproduced and still need a fix. Once an issue is assigned to a developer, its status will change to \textit{ASSIGNED}. After the resolution, it will change to \textit{RESOLVED}. However, a Quality Assurance (QA) team shall review the resolution and either change the status to \textit{VERIFIED} or reopen the bug by setting the status field to \textit{REOPEN}. Once verified, the status will change to \textit{CLOSED} \cite{bugzilla-lifecycle1, bugzilla-lifecycle2}. Furthermore, \textit{product} and \textit{component} are two vital fields showing the relevant software project (i.e., product) and the specific component in that software project that is related to the issue. Prior works have demonstrated a high predictive power for the product and component fields when used as Machine Learning (ML) features for training an ML model \cite{Zhang+2019}.

\subsection{\textbf{Natural Language Processing (NLP)}}

Natural Language Processing (NLP) has evolved from natural language understanding. NLP is an amalgamation of computational procedures, including syntactical, discourse, and semantic analysis, among many other processes, which, in turn, lead to the examination of language by machines \cite{Chowdhary+2020}. NLP is a sub-discipline of Artificial Intelligence (AI), which benefits from techniques, algorithms, and methods in other AI sub-disciplines, such as Machine Learning (ML) and Information Retrieval (IR).

While it is possible to obtain analysis from machines, the main downfall is that a machine still underachieves compared to the level of analysis of humans at this stage \cite{Chowdhary+2020}. Large Language Models (LLMs), such as the Generative Pre-trained Transformers (GPT), have enhanced the NLP performance significantly. However, they still lack a deep understanding of the text and provide no actual reasoning. Instead, they work based on co-occurrences in textual corpora. Hence, it is no surprise that they sometimes suffer from the so-called \textit{LLM hallucinations}, while responding in a confident manner. Despite this, their performance in solving real-world problems has been fascinating.

There are different ways to improve the performance of NLP methods. First, one can preprocess the data better before feeding them into NLP methods. Preprocessing can tremendously reduce the noise within a dataset, making it much simpler for a machine to process efficiently \cite{Ali+2024}. Some standard preprocessing techniques with natural language processing include word tokenization, stop-word removal, and lemmatization \cite{Tian+2015,Umer+2018,Xia+2017}. Word tokenization extracts words as tokens from a text with a delimiter for separation between tokens \cite{Tian+2015}. Stop-word removal ensures that only the more important words will be incorporated into the set of tokens, not the so-called stop-words (e.g., \textit{the}, \textit{a}, \textit{an}, etc.) that appear frequently, and lemmatization transforms tokens into their baseline form (e.g., \textit{transforming} and \textit{transformed} would both become \textit{transform}) \cite{Umer+2018}. In this paper, we conduct word tokenization and vectorization, where words are transformed into vector representations.

Another technique to enhance NLP methods is to deploy them into a pipeline of methods where they deal with a more focused set of instances that have already passed through other filters in the pipeline, or in some cases the query sent to them may be somehow engineered and refined in a manner that cab yield better results. In this work, we create a pipeline that, as briefly described in Section \ref{introduction}, comprises a topic modeling step and a text classification step.

\section{Related Work} \label{related-work}

(Semi-)Automated bug triage has been explored over the past decades. Various approaches and techniques, such as using machine learning for bug localization and automated bug assignment based on that \cite{MoinKhansari2010}, or automated bug assignment using information retrieval based on developers' profiles \cite{MoinNeumann2012}, were proposed.

In this section, we review the related works in the literature related to topic modeling, which is relevant for step 1 in our method, and automated bug report prioritization, which is related to the overall goal of this research paper.

\subsection{\textbf{Topic Modeling}}

Blei et al. proposed \textit{Latent Dirichlet Allocation (LDA)}, a three-tiered form Bayesian model used as a generative probabilistic model for a specific set of vocabulary \cite{blei+2003}. It is still used in many forms of research today. However, most papers use it as a baseline for their comparison to the state-of-the-art topic modeling approaches. 

Xia et al. improved bug triage techniques using a topic model referred to as the \textit{Multi-feature Topic Model} (MTM) within their \textit{TopicMiner} prototype \cite{Xia+2017}. MTM incorporated the product and component elements of a bug report and achieved an appropriate topic distribution. They tested the effectiveness of MTM over LDA by directly comparing the results obtained with their TopicMiner\(^{MTM}\) with that of TopicMiner\(^{LDA}\). While their features were similar, TopicMiner\(^{MTM}\) outperformed TopicMiner\(^{LDA}\).

An alternative to MTM or LDA is \textit{Neural Topic Models} (NTMs) \cite{Wu+2024}. This approach is more scalable and flexible. However, it still suffers from serious issues, for example, with respect to its reliability and the quality of its topic models \cite{Wu+2024}. 

In this paper, we deploy a similar strategy to Xia et al.'s TopicMiner\(^{MTM}\) \cite{Xia+2017} in step 1 of our method.

\subsection{\textbf{Bug Prioritization}}

Ali et al. \cite{Ali+2024} predicted the severity of bug reports using Bidirectional Encoder Representations from Transformers (BERT) \cite{Devlin+2019}, a state-of-the-art Large Language Model (LLM). They selected some publicly available datasets, computed a value for an associated sentiment for each bug report within their datasets, and deployed BERT for data formatting and creating prioritization predictions. Their results surpassed other approaches by a significant amount, partially due to their utilization of BERT. BERT is advantageous as it responds better to high-paced development and has a richer natural language understanding, a characteristic essential for text classification \cite{Ali+2024}.

Moreover, DRONE is an automated priority assessment approach presented by Tian et al. \cite{Tian+2015} to base predictions on multi-faceted factor analysis. They also created a text classification engine that enhanced linear regression. Their findings improved their previous approach \cite{Tian+2013} significantly. 

Further, Umer et al. \cite{Umer+2018} investigated the use of emotions presented within bug reports to enhance prioritization techniques to accurately determine the priority level for a specific bug report using DRONE. They created an emotional value calculation with approximately 50,000 positive and negative words based on bug report descriptions. They found a positive correlation between sentiment and bug priority and used that to outperform other approaches \cite{Umer+2018}. However, since this is an open-set machine learning problem, as anyone can report new bug reports to the repository, some new challenges may arise over time.

Additionally, Yadav and Rathore \cite{Yadav+2024} attempted to overcome existing challenges in automated bug prioritization using existing approaches, such as DRONE, by deploying Hierarchical Attention Networks (HANs). HANs leverage the structure presented within a text to analyze the content at multiple levels. They first carried out pre-processing, tokenization, and embeddings with the help of DistilBERT \cite{Sanh+2019}. Then, using the HAN model for feature extraction, final predictive values were given. Their accuracy outperformed other approaches in the literature.

\section{Proposed Approach} \label{proposed-approach}
In this work, the automated bug prioritization problem is formalized as follows: Given a set of issues (i.e., bug reports), $\{b_1, b_2, ..., b_n\}$ reported to the issue tracking system of an open-source project and their associated priority levels, which can be any of $\{P_1, P_2, ..., P_5\}$, we want to be able to predict the right priority level for a new bug report, $b_{n+1}$ automatically, based on the textual information that is contained in the bug report.

To this aim, we deploy a machine learning (ML) model and train it with the textual information and the eventual priority labels of $\{b_l, ... b_m\} \subset \{b_1, b_2, ..., b_n\}$, where $\forall x \in \{b_l, ... b_m\}$ $ status(x) == RESOLVED\_FIXED$.

The reason that we train our ML model only on bug reports that are labeled as \textit{RESOLVED\_FIXED} is that we want to make sure our classifier will see not much noise and incorrect labels as opposed to reasonable labels. For instance, a bug report's priority level may not have been assigned accurately at the beginning of its lifecycle. However, the assumption is that by its resolution time, the priority level has been adjusted if necessary.

Given a new bug report $b_{n+1}$, for which $status(b_{n+1}) == NEW$, our trained ML model will be capable of assessing its priority level $P_{n+1} \in \{P_1, P_2, ..., P_5\}$.

\begin{figure*}
    \centering
    \includegraphics[width=0.6\textwidth]{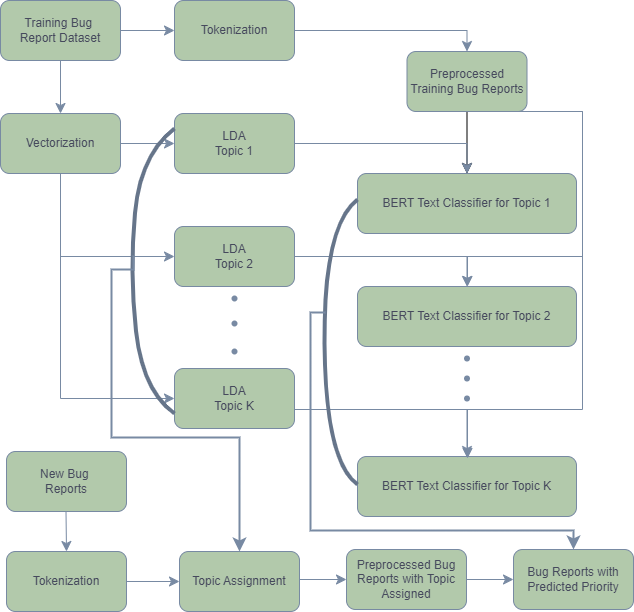}
    \caption{Our BERT Approach Overview \cite{BERT}}
    \label{fig:ApproachVisualization}
\end{figure*}

\begin{figure*}
    \centering
    \includegraphics[width=0.6\textwidth]{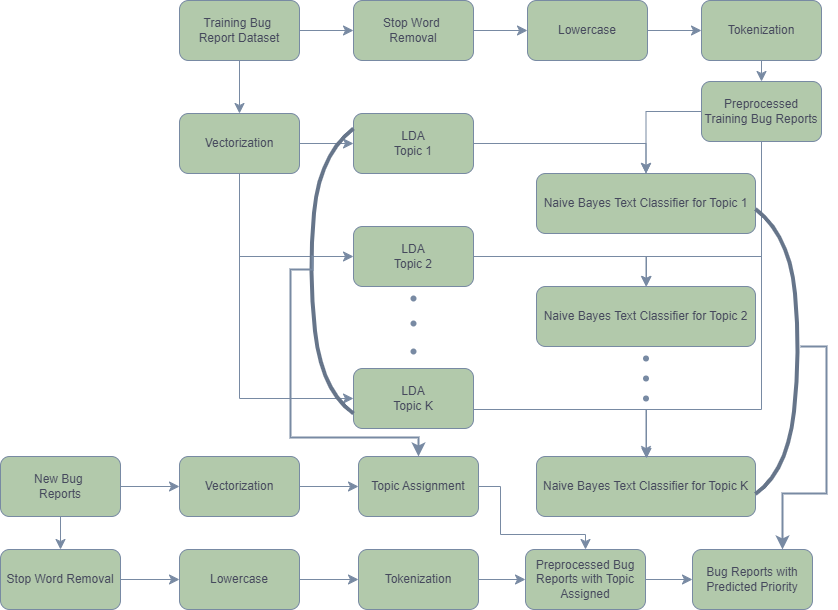}
    \caption{Our Na\"ive Bayes Approach Overview}
    \label{fig:ApproachVisualization2}
\end{figure*}

We propose a novel approach to the above-mentioned problem in large open-source projects based on BERT \cite{Devlin+2019}. Figure \ref{fig:ApproachVisualization} illustrated our overall approach. Moreover, we deploy a Na\"ive Bayes classifier to acquire some baseline results as a reference point for comparison. The latter approach is depicted in Figure \ref{fig:ApproachVisualization2}.

Our approach comprises two steps: Topic modeling and text classification. Therefore, during the training phase, bug reports are categorized into a number of topics, and then a text classifier is trained for each topic. This classifier can predict the priority level of a bug report belonging to this specific topic, given the bug report's textual information, including the summary/title, description, product, and component.

During the testing phase or when the system is deployed in production, when a new bug report arrives in the bug repository, first, the most relevant topic is determined, and then the classifier associated with that particular topic will be queried to predict the correct priority level for the new bug report. In the following, we briefly explain each step.

\subsection*{Step 1: Topic Modeling}\label{topic-model}

After vectorization, we create topic models for the bug reports. To ensure that the topics are relevant, we incorporate a similar strategy to Xia et al.'s approach, MTM \cite{Xia+2017}: We use the \textit{summary/title}, \textit{description}, \textit{product}, and \textit{component} data for topic modeling. Xia et al.'s \cite{Xia+2017} research showed that these fields possess a high predictive power. Moreover, MTM is based on the well-established LDA method for topic modeling. Since MTM's source code is not publicly available, we decided to use LDA in the topic modeling step's implementation. However, our work is certainly inspired by MTM, for example, regarding the textual data fields that should be used (as explained above). 

\subsection*{Step 2: Text Classification}\label{classification}

After topic modeling, we train multiple text classifiers using BERT, Multinomial Na\"ive Bayes, and Gaussian Na\"ive Bayes for each topic. The Na\"ive Bayes classifiers are deployed to achieve baseline results. As Figures \ref{fig:ApproachVisualization} and \ref{fig:ApproachVisualization2} show, the Na\"ive Bayes method includes some additional data pre-processing for removing the stop words and lower-casing all remaining words. This improves the prediction results. BERT does not require these steps.

The textual data fields of bug reports that are used in this step to train the mentioned machine learning (ML) models are as follows: summary/title, description, and component. Hence, the trained ML models will become capable of predicting a new bug report's priority level based on the textual data in its summary/title, description, and component.

\section{Experimental Study} \label{experimental-results}
\subsection{Research method}
Our empirical research method in this study is Repository Mining \cite{acm-empirical-methods-repo-mining}. Below, we describe our evaluation metrics, dataset, and experimental results.

\subsection{Evaluation metrics}\label{evaluation}

To evaluate the proposed approach, in the following, we compare our results, which are achieved by using the BERT model, with a baseline method, namely Na\"ive Bayes. We also compare them with related work \cite{Yadav+2024, Umer+2018, Tian+2015, Tian+2013} in Section \ref{discussion}. To this aim, we use the typical classification performance indicators, namely Accuracy, Precision, Recall, and F1-measure. Table \ref{tab:metrics} shows how these performance metrics are defined. TP, TN, FP, and FN stand for the number of True Positive, True Negative, False Positive, and False Negative cases, respectively. Since we have five classes for the five priority levels, thus no binary classification (positive and negative), we average over all classes to calculate the metrics. Therefore, $i \in \{1, 2, ..., 5\}$.

\begin{figure}[hbt!]
    \centering
    \includegraphics[width=0.75\linewidth]{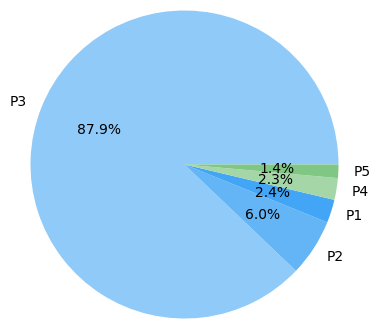}
    \caption{\centering{Priority level distribution of the Eclipse Platform project's bug reports \cite{datasethomepage,eclipsedataset}}.}
    \label{fig:PriorityLevels}
\end{figure}

Further, the main difference between \textit{micro-averaging} and \textit{macro-averaging} is how skewed data is handled. In micro-averaging, adopted by most related work in this area, class imbalance, which can be significant, has not been taken into account. This means skewed data can make classifiers considerably biased toward the more populated classes. In contrast, in macro-averaging, weights are associated to make classes balanced. Since the priorities are not evenly distributed, with a priority of P3 overtaking $87.9\%$ of the data, as shown in Figure \ref{fig:PriorityLevels}, macro-averaging sounds like a promising approach. Thus, we adopted this approach.

\begin{table*}[hbt!]
    \centering
    \caption{Typical classification performance metrics}
    \resizebox{.75\textwidth}{!}{%
    \begin{tabular}{|c|c|c|}\hline 
         Metric & Micro-Averaging & Macro-Averaging  \\ \hline 
         \rule{0pt}{15pt}
         \raisebox{.5\height}{Accuracy} & \multicolumn{2}{c|}{\raisebox{.5\height}{$\frac{\sum_i TP_i + \sum_i TN_i}{\sum_i TP_i + \sum_i FP_i + \sum_i FN_i + \sum_i TN_i}$}}\\ \hline 
         \rule{0pt}{15pt}
         \raisebox{1\height}{Precision} & \raisebox{.5\height}{$\frac{\sum_i TP_i}{\sum_i TP_i + \sum_i FP_i}$} &  \raisebox{.5\height}{$\frac{1}{n} \sum_i \frac{TP_i}{TP_i + FP_i}$} \\ \hline 
         \rule{0pt}{15pt}
         \raisebox{1\height}{Recall} & \raisebox{.5\height}{$\frac{\sum_i TP_i}{\sum_i TP_i + \sum_i FN_i}$} & \raisebox{.5\height}{$\frac{1}{n} \sum_i \frac{TP_i}{TP_i + FN_i}$} \\ \hline 
         \rule{0pt}{30pt}
         \raisebox{1.75\height}{F1-measure} & \raisebox{1.2\height}{$\frac{2 \sum_i TP_i}{2 \sum_i TP_i + \sum_i FP_i + \sum_i FN_i}$} & \raisebox{.75\height}{$\frac{1}{n} \sum_i \frac{2 \cdot \frac{TP_i}{TP_i + FP_i} \cdot \frac{TP_i}{TP_i + FN_i}}{\frac{TP_i}{TP_i + FP_i} + \frac{TP_i}{TP_i + FN_i}}$} \\ \hline 
    
    \end{tabular}%
    }
    \vspace{.1cm}
    \label{tab:metrics}
\end{table*}

\subsection{Dataset}\label{Dataset}
We use an open dataset with 85,156 bug reports with resolution dates between 2001 and 2014. These bugs are much less likely to change priority status after ten or more years \cite{Tian+2015}. The issues belong to the Eclipse Platform open bug repository \cite{eclipsedataset}. 

\subsection{Step 1: Latent Dirichlet Allocation (LDA) Topic Modeling}

\begin{figure}[hbt!]
    \centering
    \includegraphics[width=1\linewidth]{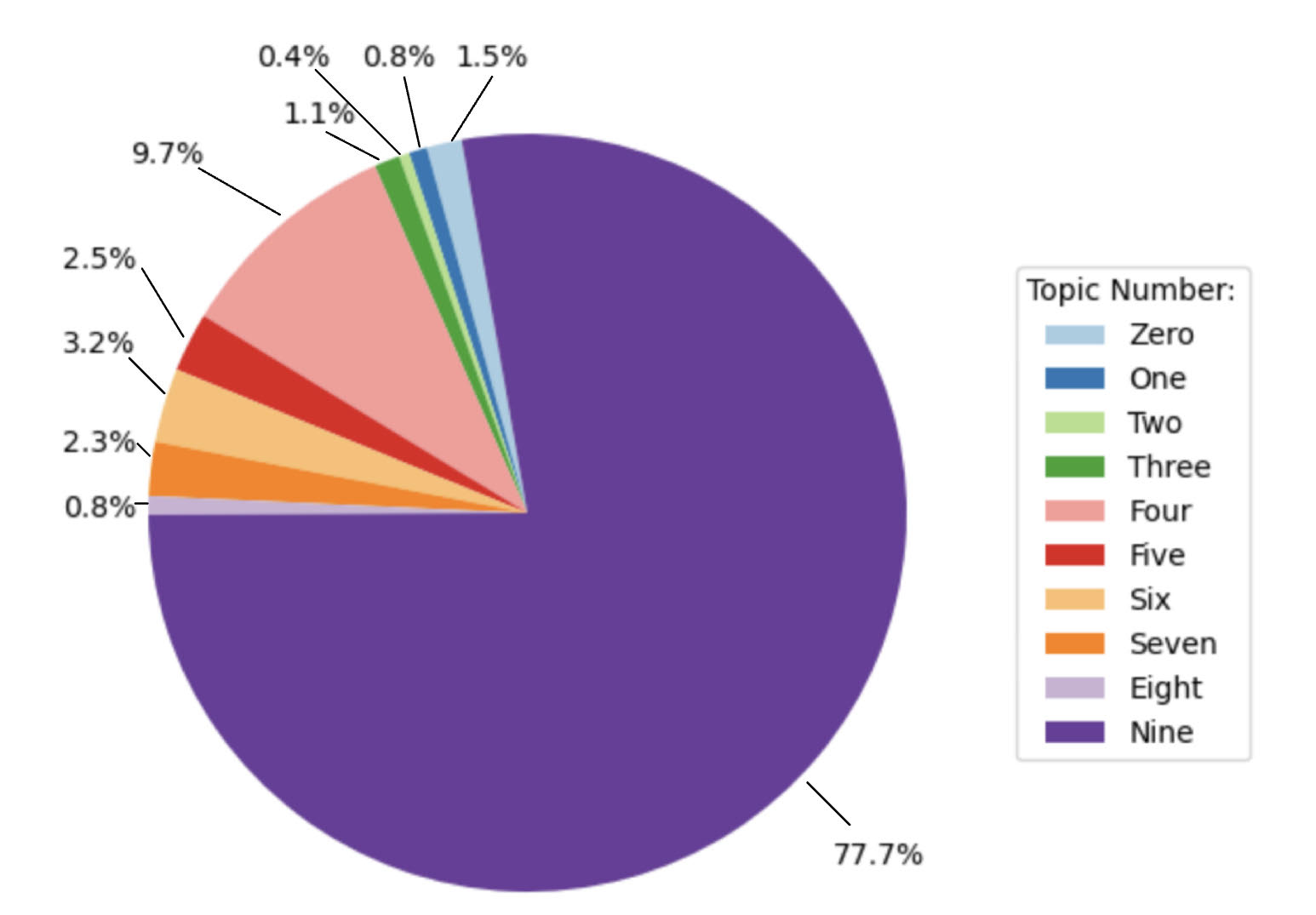}
    \caption{Topic Distributions Using LDA}
    \label{fig:LDA Topics}
\end{figure}

We used the summary/title, component, and description of bug reports with the Latent Dirichlet Allocation (LDA) method \cite{blei+2003} to create the topics and find the most relevant topic for a new bug report. According to the experimental results, the component field has a high predictive power. The LDA method created ten topics with varying numbers of bug reports assigned to each topic. As Figure \ref{fig:LDA Topics} illustrates, one of the topics (topic number nine) has overtaken most of the data.

\subsection{Step 2a: The BERT Text Classification}\label{BERTResults}

BERT has shown proven text classification performance in prior works \cite{Ali+2024, Yadav+2024}, as well as in our experimental study. In this part, we report on our experimental results with BERT. However, we also note that it took excessive time to train the model compared to the baseline approaches, such as Na\"ive Bayes, for which the results will be reported in the following part. Regarding the testing phase, BERT also took much longer to classify bug reports than the baseline methods. The training and testing times are reported in Tables \ref{tab:training-times} and \ref{tab:testing-times}, respectively. 

When we trained BERT for text classification, we used 15 epochs for topics 0 to 8. However, for topic number 9, since it takes up most of the bug reports as shown in Figure \ref{fig:LDA Topics}, it would take more time than the other topics combined. Therefore, when training our model on this topic, we did it with 1 and also with 3 epochs instead of 15. Table \ref{tab:epochs} shows the results of using 1 versus 3 epochs for this topic. Interestingly, 1 epoch yielded better results than 3 epochs. Therefore, when comparing our pipeline's results using BERT to baseline approaches, we rely on our results from 1 epoch for topic nine.

\subsection{Step 2b: The Na\"ive Bayes Text Classification}\label{Naive_Bayes_Results}

To evaluate our approach, we compared our BERT model results to two baseline methods using Na\"ive Bayes (NB) for text classification (see Table \ref{tab:results}) in addition to comparisons with state-of-the-art methods (see Table \ref{tab:results-compared-sota}). 

All of our pipelines use LDA to create topic models (i.e., for step 1), though they deviate when it comes to their text classifiers (i.e., step 2). For the first comparison to BERT, we use a Gaussian Na\"ive Bayes classifier for each topic to predict the bug priorities. Also, we use a Multinomial Na\"ive Bayes classifier for the second comparison. 

The Multinomial Na\"ive Bayes method had slightly better results within the micro-analysis than the Gaussian Na\"ive Bayes method. However, when performing a macro-analysis, Gaussian Na\"ive Bayes performed marginally better in precision and recall as it could more easily handle the imbalanced priority levels within our dataset. One of the main issues with the Multinomial Na\"ive Bayes classifier is that it assigns priority level P3 for almost every bug report. Note that P3 is very common in the dataset.

\begin{table*}[hbt!]\label{Efficiency_Table_Train}
    \centering
    \caption{\centering{Training Time and Memory Usage}}
    \resizebox{.75\textwidth}{!}{%
    \begin{tabular}{|c|c|c|} \hline 
          \textbf{Priority} & \textbf{Time} & \textbf{Peak Memory} \\ 
          \textbf{Method} & \textbf{Approximation} & \textbf{Usage Estimation} \\ \hline
          Gaussian NB Training & 7.22 Minutes & 10393 MB  \\ \hline 
          Multinomial NB Training& 3.46 Minutes & 4075 MB \\ \hline 
          BERT Training & 12+ Hours & 11264+ MB \\ \hline
    \end{tabular}%
    }
    \vspace{.1cm}
    \label{tab:training-times}
\end{table*}

We trained our classifiers with 68,124 bug reports (i.e., $80\%$ of the entire data). The Multinomial Na\"ive Bayes classifier took the shortest amount of time and the least memory. Table \ref{tab:training-times} shows that the Multinomial Na\"ive Bayes classifier takes less than half of the time of the Gaussian Na\"ive Bayes model and a small fraction of the time compared to the BERT model's training time. Note that our BERT model's peak memory usage exceeded the capacity of our GPU. Therefore, we partially relied on the support of our CPU during training.

\begin{table*}[hbt!]\label{Efficiency_Table_Test}
    \centering
    \caption{\centering{Testing Time and Memory Usage}}
    \resizebox{.75\textwidth}{!}{%
    \begin{tabular}{|c|c|c|} \hline 
          \textbf{Priority} & \textbf{Time} & \textbf{Peak Memory} \\ 
          \textbf{Method} & \textbf{Approximation} & \textbf{Usage Estimation} \\ \hline
          Gaussian NB Testing & 5.80 Minutes & 144 MB  \\ \hline 
          Multinomial NB Testing& 8.61 Minutes & 144 MB \\ \hline 
          BERT Testing & 76.82 Minutes & 482 MB \\ \hline 
    \end{tabular}%
    }
    \vspace{.1cm}
    \label{tab:testing-times}
\end{table*}

\begin{table*}[hbt!]\label{BERT_Epochs}
    \centering
    \caption{\centering{BERT Results Using Different Epochs for Topic Nine, Micro-Averaging}}
    \resizebox{.75\textwidth}{!}{%
    \begin{tabular}{|c|c|c|c|c|} \hline 
          \textbf{Epochs} & \textbf{Precision} & \textbf{Recall} & \textbf{F-Measure} & \textbf{Accuracy}\\ \hline
          3 & 0.6087 & 0.5895 & 0.5989 & 0.9676 \\ \hline 
          1 & 0.6579 & 0.6514 & 0.6546 & 0.9667 \\ \hline 
    \end{tabular}%
    }
    \vspace{.1cm}
    \label{tab:epochs}
\end{table*}

\begin{table*}[hbt!]\label{Analysis_Table}
    \centering
    \caption{\centering{Comparison of Performance Metrics for Our Approaches}}
    \resizebox{1\textwidth}{!}{%
    \begin{tabular}{|c|c|c|c|c|c|} \hline 
         \textbf{Analysis Method} & \textbf{Prioritization Method} & \textbf{Precision} & \textbf{Recall} & \textbf{F-Measure} & \textbf{Accuracy}\\ \hline 
         \multirow{2}{*}{Micro} & Gaussian & 0.4784 & 0.5234 & 0.4999 & 0.7517 \\ \cline{2-6}
                                & Multinomial & 0.4935 & 0.4968 & 0.4951 & \textbf{0.9743}\\ \cline{2-6}
                                & BERT & \textbf{0.6579} & \textbf{0.6514} & \textbf{0.6546} & 0.9667\\ \hline 
                                \hline
         \multirow{2}{*}{Macro} & Gaussian & 0.2020 & \textbf{0.2203} & 0.1839 & 0.7517 \\ \cline{2-6}
                                & Multinomial & 0.1949 & 0.2000 & 0.1974 & \textbf{0.9743}\\ \cline{2-6}
                                & BERT & \textbf{0.3998} & 0.2079 & \textbf{0.2140} & 0.9667\\ \hline
    \end{tabular}%
    }
    \vspace{.1cm}
    \label{tab:results}
\end{table*}

\begin{table*}[hbt!]\label{Comparison_Table}
    \centering
    \caption{\centering{Micro-Averaging Performance Metrics for Priority Assessment Methods}}
    \resizebox{.85\textwidth}{!}{%
    \begin{tabular}{|c|c|c|c|c|} \hline 
          & \textbf{Priority Method} & \textbf{Precision} & \textbf{Recall} & \textbf{F-Measure}\\ \hline 
         \multirow{2}{*}{Our Methods} & Gaussian NB & 0.4784 & 0.5234 & 0.4999  \\ 
                                    &Multinomial NB & 0.4935 & 0.4968 & 0.4951\\ 
          &BERT & \textbf{0.6579} & \textbf{0.6514} & \textbf{0.6546}\\ \hline
          \multirow{2}{*}{Baseline Methods} &                       Emotion Based \cite{Umer+2018} & 0.5762 & 0.3878 & 0.4636\\ 
                                 &DRONE \cite{Tian+2015} & 0.2566 & 0.2669 & 0.2594\\ 
                                 &DRONE \cite{Tian+2013} & 0.2973 & 0.3202 & 0.2974\\ 
                                 &HAN \cite{Yadav+2024} & 0.3952 & 0.4002 & 0.3739\\ \hline 
    \end{tabular}%
    }
    \label{tab:results-compared-sota}
\end{table*}

After training the various classifiers, we tested them using the remaining 17,032 bug reports (i.e., $20\%$ of the entire data). Note that we did not shuffle the data so we can preserve the order of instances, as bug reports are inherently sequential (time series) data. For the Gaussian Na\"ive Bayes classifier, it took 0.02 seconds to prioritize a single bug report. Also, a bug report was prioritized in 0.03 seconds with the Multinomial Na\"ive Bayes classifier, and in 0.27 seconds using the BERT classifier.

Once the Na\"ive Bayes models are trained, contrary to their respective memory usage during training, they use the same amount of memory to classify bug reports. While the Gaussian Na\"ive Bayes classifier took longer in the initial training phase than the Multinomial Na\"ive Bayes classifier, in testing, the Gaussian Na\"ive Bayes model took merely two-thirds the time of the Multinomial Na\"ive Bayes model. BERT took much more time and memory than the Na\"ive Bayes alternatives. However, it took about a quarter of a second to classify a bug report and less than half of a Gigabyte (GB) of memory with BERT. 

\section{Discussion} \label{discussion}

\subsection{Comparing to baseline methods}
In Table \ref{tab:results}, we compare our results with the Gaussian Na\"ive Bayes and Multinomial Na\"ive Bayes methods. Our Precision surpassed these baseline methods by 8.17 percent, Recall by 25.12 percent, and F1-measure by 19.1 percent. The downfall of our approach, however, is the amount of time it takes BERT to train and predict compared to the other methods. 

The Gaussian Na\"ive Bayes model achieved higher results than any of the other methods in terms of Recall. The likely cause of its higher Recall score compared to Precision is that there may be more falsely labeled priorities than seen within Umer et al.'s research \cite{Umer+2018}. The Gaussian Na\"ive Bayes method also superseded other approaches in terms of F1-measure. While our Precision may not exceed others', our method's F1-measure superiority is promising, given the typical trade-off between Precision and Recall.

\subsection{Comparing to the state of the art}
In Table \ref{tab:results-compared-sota}, we compare our results using the Gaussian Na\"ive Bayes, Multinomial Na\"ive Bayes, and BERT with the works of Yadav et al. \cite{Yadav+2024}, Umer et al. \cite{Umer+2018}, and Tian et al. \cite{Tian+2015, Tian+2013}. It is evident that our BERT model outperforms the state of the art. The comparison is based on the results reported in those papers. Also, the reason that Accuracy is missing in this table is that the other papers had not reported their Accuracy values in the papers.

We deployed LDA \cite{blei+2003} for our topic modeling. Similar to MTM, we incorporated more natural language text than simply the description \cite{blei+2003, Xia+2017}. Moreover, we differed in our topic modeling approach from Wu et al. \cite{Wu+2024} because the NTM approach is not as reliable as LDA.

Regarding text classification, our most effective method used a BERT model for each topic, similar to Ali et al. \cite{Ali+2024}. However, it is worth noting that Ali et al. \cite{Ali+2024} concentrated on the task of severity prediction.

Tian et al. \cite{Tian+2013, Tian+2015} created their text classification engine called DRONE. Our experimental results show that BERT outperforms DRONE in the bug prioritization task.

\subsection{Threats to Validity} \label{threats-to-validity}
Our work and the related work papers, with which we compared our experimental results, have used various datasets. Therefore, although we have compared the achieved results for various metrics, in principle, since the datasets are not identical, the comparison may not be very accurate. Further, the achieved results might be specific to the selected dataset. More research will be required to study the performance of the methods on other datasets.

Further, our work has focused on Bugzilla with the five priority levels, while not all open-source software projects use Bugzilla. In fact, many other choices, such as GitHub issues, do not even have a notion of importance or priority for bug reports. 

\section{Conclusion and Future Work} \label{conclusion-future-work}
In this paper, we have proposed a novel approach to automated bug prioritization, an important part of the bug triage process in large open-source projects. We deployed the BERT model and used the textual natural language information in bug reports to assess their importance/priority. Experimental results have shown that we outperform the state-of-the-art and some baseline approaches.

In our future work, we will seek improvement in several areas. First, we plan to use the same dataset and similar experimental environments to benchmark the related work and our methods. Second, we want to enable sentiment analysis, similar to Umer et al. \cite{Umer+2018}, to enhance the results. Our hypothesis is that sentiments of bug reports should help increase the performance of automated bug prioritization. Further, we used a very basic vectorization method. More sophisticated methods, such as TF-IDF and Word2Vec, should be explored. Finally, we want to deploy the GPT models and see how these models perform compared to the BERT large language model.

\section*{Software and Data Availability}
The prototype is available under a permissive open-source license at \url{https://github.com/qas-lab/PiersonREU}. The data used for evaluation are also publicly available at \cite{eclipsedataset}.

\section*{Acknowledgment}
This material is based upon work supported by the U.S. National Science Foundation (NSF) under Grant No. 2349452. Any opinions, findings, conclusions, or recommendations expressed in this material are those of the authors and do not necessarily reflect the views of the NSF.

\bibliography{refs}
\bibliographystyle{IEEEtran}

\end{document}